\documentclass{PoS}
\usepackage[utf8]{inputenc}
\usepackage{booktabs}
\usepackage{amsmath}
\newcommand{\msbar}{{\rm \overline{MS\kern-0.05em}\kern0.05em}}

\title{Comparison between models of QCD with and without dynamical charm quarks}

\ShortTitle{Comparison between models with and without dynamical charm quarks}
\author{\speaker{Salvatore Cal\`i}\\
        Department of Physics, Bergische Universit\"at Wuppertal, Gau\ss str. 20, 42119 Wuppertal, Germany\\
        Department of Physics, University of Cyprus, P.O. Box 20537, 1678 Nicosia, Cyprus\\
        E-mail: \email{scali@uni-wuppertal.de}}
        
\author{Francesco Knechtli\\
        Department of Physics, Bergische Universit\"at Wuppertal, Gau\ss str. 20, 42119 Wuppertal, Germany\\
        E-mail: \email{knechtli@physik.uni-wuppertal.de}}
        
\author{Tomasz Korzec\\
        Department of Physics, Bergische Universit\"at Wuppertal, Gau\ss str. 20, 42119 Wuppertal, Germany\\
        E-mail: \email{korzec@uni-wuppertal.de}}       

\abstract{We investigate the influence of dynamical charm quarks on observables that depend explicitly on the charm quark fields, like the pseudo-scalar and vector masses $m_{\eta_c}$ and $m_{J/\psi}$, the hyperfine splitting 
$(m_{J/\psi}-m_{\eta_c})/m_{\eta_c}$, the charm quark mass and the meson decay constants.
For this purpose, instead of working in full QCD we study a simplified setup. We simulate two theories: $N_f= 0$ QCD and QCD with $N_f= 2$ dynamical quarks at the charm mass. The absence of light quarks allows us to reach extremely fine lattice spacings (0.02 fm < a < 0.05 fm) which are crucial for reliable continuum extrapolations. Our main result is a comparison of various quantities in the continuum limit. For the hyperfine splitting we find that the effects of a dynamical charm quark are below our statistical precision of $2\%$.}

\FullConference{The 36th Annual International Symposium on Lattice Field Theory - LATTICE2018\\
		22-28 July, 2018\\
		Michigan State University, East Lansing, Michigan, USA.}

\begin{document}

\section{Introduction}
In this work we want to study the charm sea effects on charmonium systems, that are bound states made of a charm quark ($c$) and a charm antiquark ($\bar{c}$). The high mass of a charm quark allows the description of $c\bar{c}$ states in terms of certain non-relativistic potential models and this makes the charmonium system an important testing ground for a comparison of theory with experiment. Moreover, in the last few years, experiments also discovered a large number of unexpected charmonium-like states, many of which are still poorly understood. This highlights the need for a more complete theoretical understanding of these systems starting from first principles and lattice QCD represents one of the most suitable tools to reach these purposes. 

Up to date many simulations of QCD are carried out using $N_f = 2 + 1$
dynamical light quarks (up, down, strange). A more complete setup would include a dynamical charm quark, but this increases the computational cost of the simulations. Thus, to give a reliable estimate of the effects of a dynamical charm quark in QCD, we compare results for physical observables obtained with two different models: $N_f=0$ QCD and QCD with $N_f=2$ degenerate charm quarks.

The absence of light quarks allows us to keep the volumes moderately large, which in turn makes simulations at extremely fine lattice spacings feasible. The main goal of this work is to determine the impact of a dynamical charm quark on meson masses and decay constants, with special focus on the mesons $\eta_c$ and $J/\psi$ (pseudo-scalar and vector channel respectively). Our first results in this direction can be found in \cite{Korzec:2016eko}. Here, we show an update of our previous results, which includes a better strategy of the tuning of the twisted mass parameter $\mu$, continuum extrapolations performed with a larger number of lattice spacings and increased statistics and our first results on the meson decay constants.
  
\section{Numerical setup}
We employ six $N_f=2$ ensembles at the charm mass $M_c$ and four $N_f=0$ ensembles, exploring a set of lattice spacings in the range $0.02\text{ fm}\lesssim a \lesssim 0.07\text{ fm}$ (see Table \ref{tab:ens}). The aim is to disentangle the charm sea effects from possible cut-off effects due to the small correlation lengths which are associated with charmonium states. As a lattice discretization scheme we use a clover improved doublet of twisted-mass Wilson fermions \cite{Sheikholeslami:1985ij, Frezzotti:2000nk, Frezzotti:2003ni} and Wilson's plaquette gauge action \cite{Wilson:1974sk} for the gluon sector. To achieve maximal twist, the hopping parameter $\kappa$ has been set to its critical value by interpolating the data published in Refs.~\cite{Fritzsch:2012wq,Fritzsch:2015eka}. Open boundary conditions in the temporal direction are imposed to keep auto-correlation times associated with the topological charge manageable \cite{Luscher:2011kk}. For further details regarding the generation of these ensembles we refer to Ref.~\cite{Knechtli:2017xgy}.

\begin{table}[ht]
\centering
{\small
\begin{tabular}{c | c c c c c c c c}
\toprule
$N_f$ & $\frac{T}{a}\times\left(\frac{L}{a}\right)^3$ &  $\beta$  & $a[\mbox{fm}]$  & $\kappa$    & $a \mu$            & $\sqrt{t_0}m_{\eta_c}$ & $t_0/a^2$ &  MDUs\\
\midrule
2 & $96\times 24^3$  &  5.300  &$0.066$  & 0.135943    & 0.36151  & 1.79321(53)      & 1.23950(85) & 8000\\
 & $120\times 32^3$                              &  5.500  &$0.049$  & 0.136638    &  0.165997          & 1.8049(16)      & 4.4730(95) & 8000\\
 & $192\times 48^3$  &  5.600  &$0.042$  & 0.136710    & 0.130949  & 1.7655(15) & 6.609(15) & 8000\\
 & $120\times 32^3$                              &  5.700  &$0.036$  & 0.136698    & 0.113200           & 1.7931(28)     & 9.104(36) & 17184\\
 & $192\times48^3$                               &  5.880  &$0.028$ & 0.136509    & 0.087626           & 1.8129(29)     & 15.622(62) & 23088\\
 & $192\times 48^3$                              &  6.000  &$0.023$ & 0.136335    & 0.072557           & 1.8075(42)     &22.39(12) &  22400\\
\midrule
0 & $120\times 32^3$                               &  6.100  &$0.049$ &    --       &    --              & --  & 4.4329(32) & 64000 \\
 & $120\times 32^3$                               & 6.340  &$0.036$ &    --       &    --              & --  & 9.034(29) & 20080\\
 & $192\times 48^3$                               &  6.672  &$0.023$ &    --       &    --              & --  & 21.924(81) & 73920\\
 & $192\times 64^3$                               & 6.900  &$0.017$ &    --       &    --              & --  & 39.41(15) & 160200\\
\bottomrule
\end{tabular}
}
\caption{Simulation parameters of our ensembles. The columns show the lattice sizes, the gauge coupling $\beta=6/{g_0^2}$, the lattice spacing in fm (determined from the $N_f=2$ scale $L_1$ \cite{Athenodorou:2018wpk} and using decoupling for $N_f=0$), the critical hopping parameter, the twisted mass parameter $\mu$, the pseudo-scalar mass in $t_0$ units, the hadronic scale $t_0/a^2$ defined in \cite{Luscher:2010iy} and the total statistics in molecular dynamics units.
}\label{tab:ens}
\end{table}

\section{Methodology}
\subsection{Meson masses and decay constants}
To extract meson masses and decay constants we compute the two-point correlation function
\begin{equation}
f_{O_{\Gamma_1},O_{\Gamma_2}}(x_0,y_0) = \frac{a^6}{L^3}\sum_{\mathbf{x},\mathbf{y}}\langle O_{\Gamma_1}(x_0,\mathbf{x})O_{\Gamma_2}^{\dagger}(y_0,\mathbf{y})\rangle,\quad O_{\Gamma_1,\Gamma_2}\in\lbrace\bar{c}_1\gamma_5c_2, \bar{c}_1\gamma_{i=1,2,3}\gamma_5 c_2, \bar{c}_1\gamma_0\gamma_5c_2\rbrace
\label{eq:2point_function}
\end{equation}
between two meson states, created and annihilated by the operators $O_{\Gamma_2}^{\dagger}$ and $O_{\Gamma_1}$ at the source and sink, having coordinates $(y_0 ,\mathbf{y})$, $(x_0 ,\mathbf{x})$ respectively. In Eq.~\eqref{eq:2point_function} $c_1$ and $c_2$ denote two flavors in a twisted mass doublet, the sum over spatial coordinates
$\mathbf{x}$, $\mathbf{y}$ is performed to compute the two-point function of the hadron at zero-momentum and the triangular brackets $\langle\cdots\rangle$ represent the expectation value of the observable on the ensemble of gauge and fermion fields. The ground state energy $am_O$ in a given channel $O$ is then determined by the weighted plateau average of the effective mass
\begin{equation}
am^{eff}_O(x_0+a/2,y_0) = \log\left( \frac{f_{OO}(x_0,y_0)}{f_{OO}(x_0+a,y_0)} \right).
\end{equation}
From the correlators \eqref{eq:2point_function} it is also possible to extract the pseudo-scalar and vector decay constants of the mesons $\eta_c$ and $J/\psi$, whose definitions in twisted mass QCD are given by \cite{Jansen:2003ir,Jansen:2009hr}
\begin{equation}
f_{\eta_c}m^2_{\eta_c} = 2\mu \langle 0\vert \bar{c}_1\gamma_5c_2\vert\eta_c\rangle,\quad f_{J/\psi}m_{J/\psi} = \frac{1}{3}\sum_{i=1}^3\langle 0\vert \bar{c}_1\gamma_i\gamma_5c_2\vert J/\psi\rangle.
\end{equation}
The twisted mass formulation of QCD is a particularly convenient setup for the pseudo-scalar decay constant $f_{\eta_c}$, because the renormalization factors $Z_P$ and $Z_{\mu}$ of $\bar{c}_1\gamma_5c_2$ and $\mu$ respectively obey $Z_PZ_{\mu} = 1$. Therefore the calculation of $f_{\eta_c}$ does not need any renormalization factors \cite{Jansen:2003ir}. As concerns the lattice calculation of $f_{J/\psi}$, the relevant matrix element must be multiplied by the renormalization factor $Z_A$ of $\bar{c}_1\gamma_i\gamma_5c_2$, which is known from Refs.~\cite{Luscher:1996jn,DellaMorte:2005xgj,DallaBrida:2018tpn} for the ensembles considered here. Since we use open boundary conditions in the temporal direction, computing the meson decay constant through an exponential fit to the two-point correlation function may lead to unreliable
results, because of the boundary effects. For this reason, we use the method described in Ref.~\cite{Bruno:2016plf}, whose advantage is to remove the unwanted boundary effects from our lattice calculation by forming a suitable ratio of the two-point correlation functions \eqref{eq:2point_function}, which contains a boundary-boundary correlator. However, when source and sink are close to the boundaries, a determination of the correlator \eqref{eq:2point_function} at a good accuracy is very difficult to achieve, because the relative precision of the solution of the Dirac equation deteriorates at large distances and this becomes much more prominent for heavy quark masses. To overcome this problem, we use the distance preconditioning for the Dirac operator proposed in Refs.~\cite{deDivitiis:2010ya,Collins:2017iud}. This procedure improves the quality of the signal at large time separations, at the price that more iterations are required for the solver to converge to the exact solution of the Dirac equation.

\subsection{Mass shifts}
To match $N_f=0$ and $N_f=2$ QCD we choose the low energy observable $m^{had}=1/\sqrt{t_0}$, where $t_0$ is the hadronic scale that can be extracted from the Wilson flow and introduced in Ref.~\cite{Luscher:2010iy}. For this observable decoupling of heavy quarks applies \cite{Weinberg:1980wa} and we can assume $\left[\sqrt{t_0}\right]^{N_f=2}_{M=M_c}=\left[\sqrt{t_0}\right]^{N_f=0}$. Our original strategy to compare $N_f = 2$ with $N_f = 0$ results is explained in \cite{Korzec:2016eko}. It relies on first performing the simulations at fixed charm mass and taking the continuum limit of the pseudo-scalar mass $\left[\sqrt{t_0}m_{\eta_c}\right]^{N_f=2}_{\text{cont.}}$ in the $N_f=2$ theory. This value is then used to set the twisted mass parameter
$\mu^{\star}$ for the measurements of meson correlators on the quenched ensembles, by matching $\left[\sqrt{t_0}m_{\eta_c}\right]^{N_f=0}=
\left[\sqrt{t_0}m_{\eta_c}\right]^{N_f=2}_{\text{cont.}}$. However, doing so there is an uncertainty in the value of $\mu^{\star}$ originating from the statistical error associated to $\left[\sqrt{t_0}m_{\eta_c}\right]^{N_f=2}_{\text{cont.}}$. Here, to refine our strategy, we compare $N_f=0$ and $N_f=2$ QCD fixing for both theories the charm mass $M_c$ such that $\sqrt{t_0}m_{\eta_c}$ approximately corresponds to its physical value, i.e. $\sqrt{t_0}m_{\eta_c}\equiv 1.8075$\footnote{Note that for $N_f=2$ QCD we find $\sqrt{t_0(M_c)}\simeq 0.11$ fm, which significantly deviates from its physical value \cite{Bruno:2016plf}.}. In $N_f=2$ QCD this requires the shift of $\mu$ and of a generic observable $R$ by Taylor expansions, for which we need to compute the derivative $dR/d\mu$. For primary observables $O_i$, such derivatives are given by
\begin{equation}
\frac{d\langle O_i\rangle}{d\mu} = - \left\langle\frac{dS}{d\mu}O_i\right\rangle + \left\langle\frac{dS}{d\mu}\right\rangle\left\langle O_i\right\rangle + \left\langle\frac{dO_i}{d\mu}\right\rangle,
\label{eq:twisted_mass_derivative1}
\end{equation}
where $S$ is the action of the system. By denoting with $D_{c_1}^{-1}(x,y)$ the inverse of the twisted mass Dirac operator for the quark flavor $c_1$, the action derivative reads (cf.\cite{Jansen:2008wv})
\begin{equation}
\left\langle\frac{dS}{d\mu}\right\rangle = -2\mu \sum_{x,y}\left\langle \text{Tr}\left[ D_{c_1}^{-1^\dagger}(x,y) D_{c_1}^{-1}(x,y)\right] \right\rangle^{\text{gauge}}.
\end{equation}
In the case of the meson correlators \eqref{eq:2point_function}, the first term of \eqref{eq:twisted_mass_derivative1} requires the computation of new Wick contractions, whilst the third term is zero because there is no explicit dependence on $\mu$. However, in this work we focus on observables $R$ (like the vector mass $\sqrt{t_0}m_{J/\psi}$, $\sqrt{t_0}M_c$, the hyperfine splitting $(m_{J/\psi} - m_{\eta_c})/m_{\eta_c}$, etc.), which in general are non-linear functions of $N_{obs}$ \textit{primary observables} (the meson correlators). Therefore, their derivatives assume the form
\begin{equation}
\frac{dR(\langle O_1\rangle,\langle O_2\rangle,\cdots\langle O_{N_{obs}}\rangle)}{d\mu} = \sum_{i=1}^{N_{obs}}\frac{\partial R}{\partial\langle O_i\rangle}\frac{d\langle O_i\rangle}{d\mu} + \frac{\partial R}{\partial \mu}.
\end{equation}

In a pure gauge theory the action does not depend on the quark masses and we need the twisted mass parameter $\mu$ only for the inversion of the Dirac operator. Thus, to reproduce the tuning value $\mu^{\star}$ for our $N_f=0$ ensembles, we carry out the measurements at three different values of the twisted mass parameter $\mu$ and the tuning point $\mu^{\star}$ is found through a linear interpolation of the measurements. The critical hopping parameters have been obtained through an interpolation of the values reported in Ref.~\cite{Luscher:1996ug}.

\section{Results}
\begin{figure}
\includegraphics[width=0.49\textwidth]{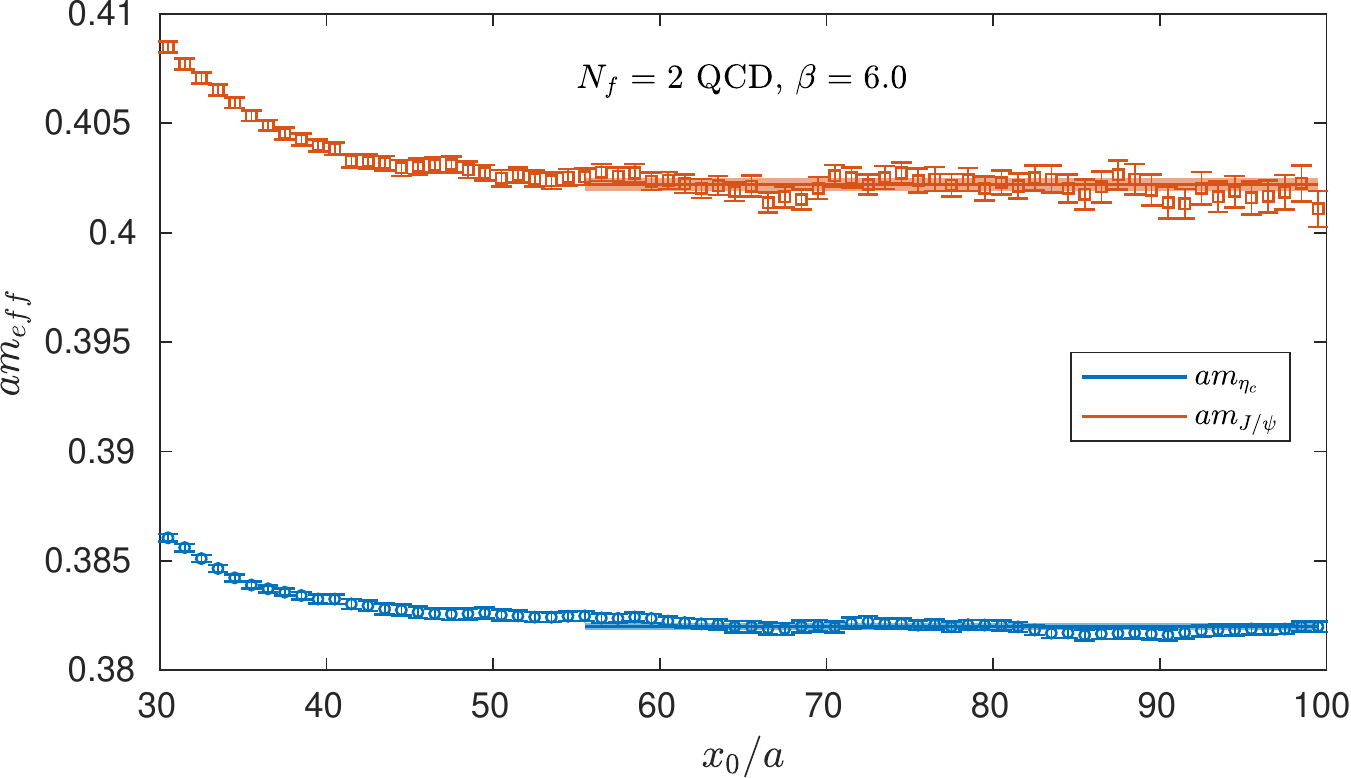}\hfill
\includegraphics[width=0.49\textwidth]{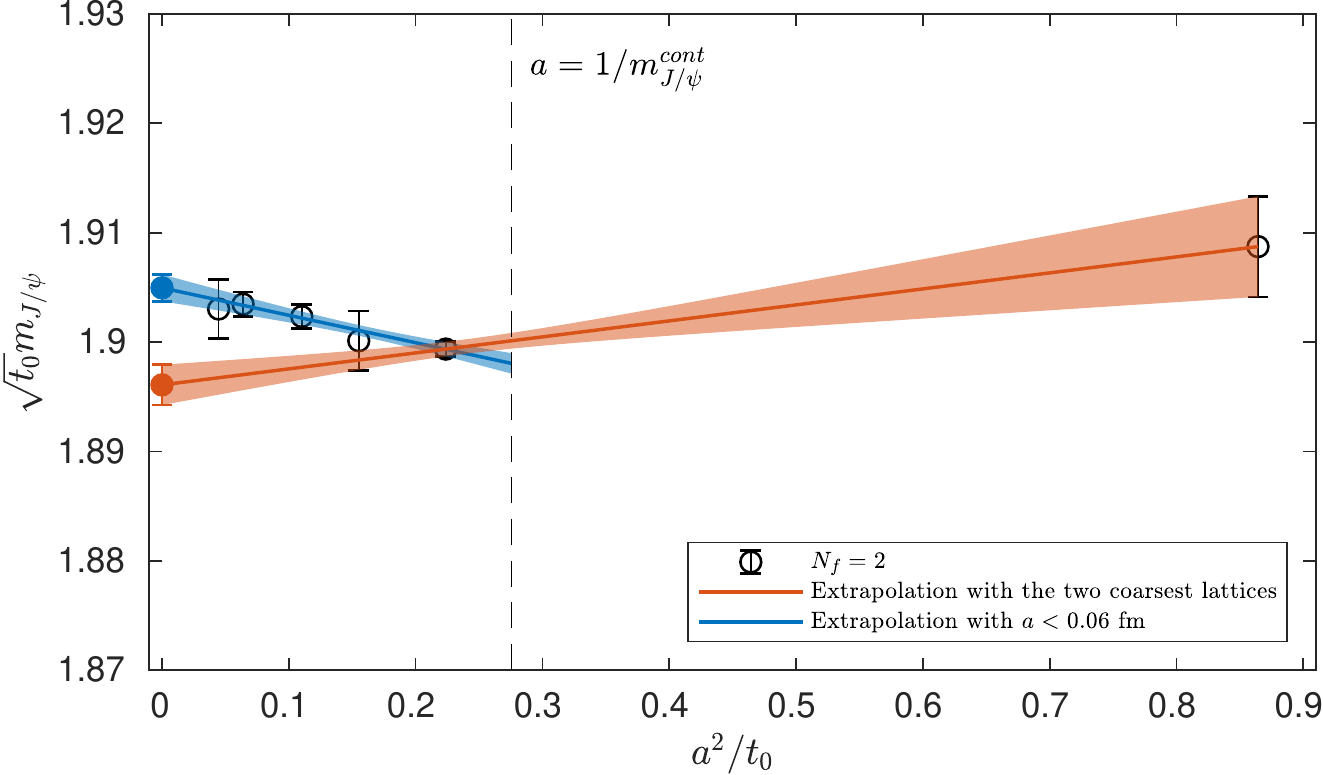}
\caption{\textit{Left panel}: effective masses and plateaux averages for the mesons $\eta_c$ (circles) and ${J/\psi}$ (squares) on the $N_f=2$ ensemble at $\beta=6.0$. \textit{Right panel}: Continuum extrapolations linear in $a^2$ of $m_{J/\psi}$ using the two coarsest lattices (red
band) and the five finest lattices (blue band) listed in the first six rows of Table 1.}
\label{fig:eff_mass_and_latt_art}
\end{figure}

\begin{figure}
\includegraphics[width=0.49\textwidth]{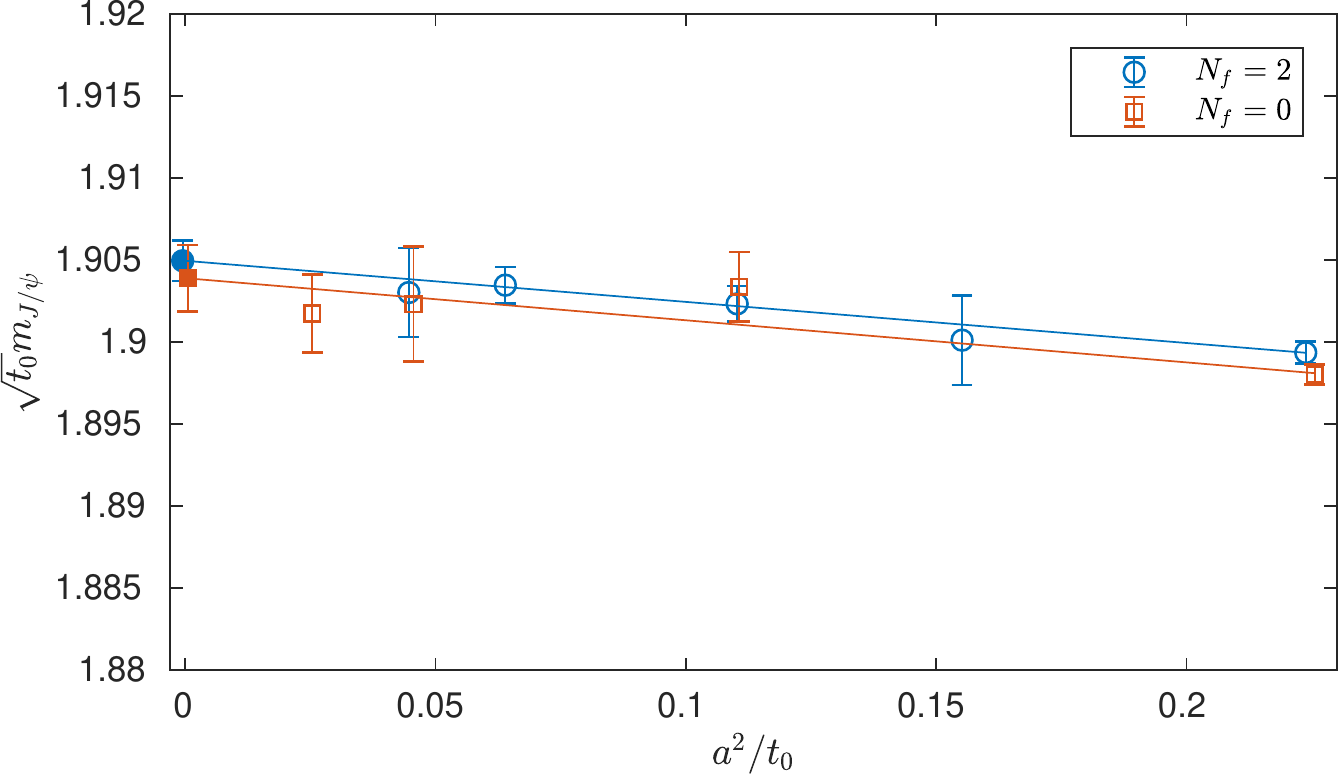}\hfill
\includegraphics[width=0.49\textwidth]{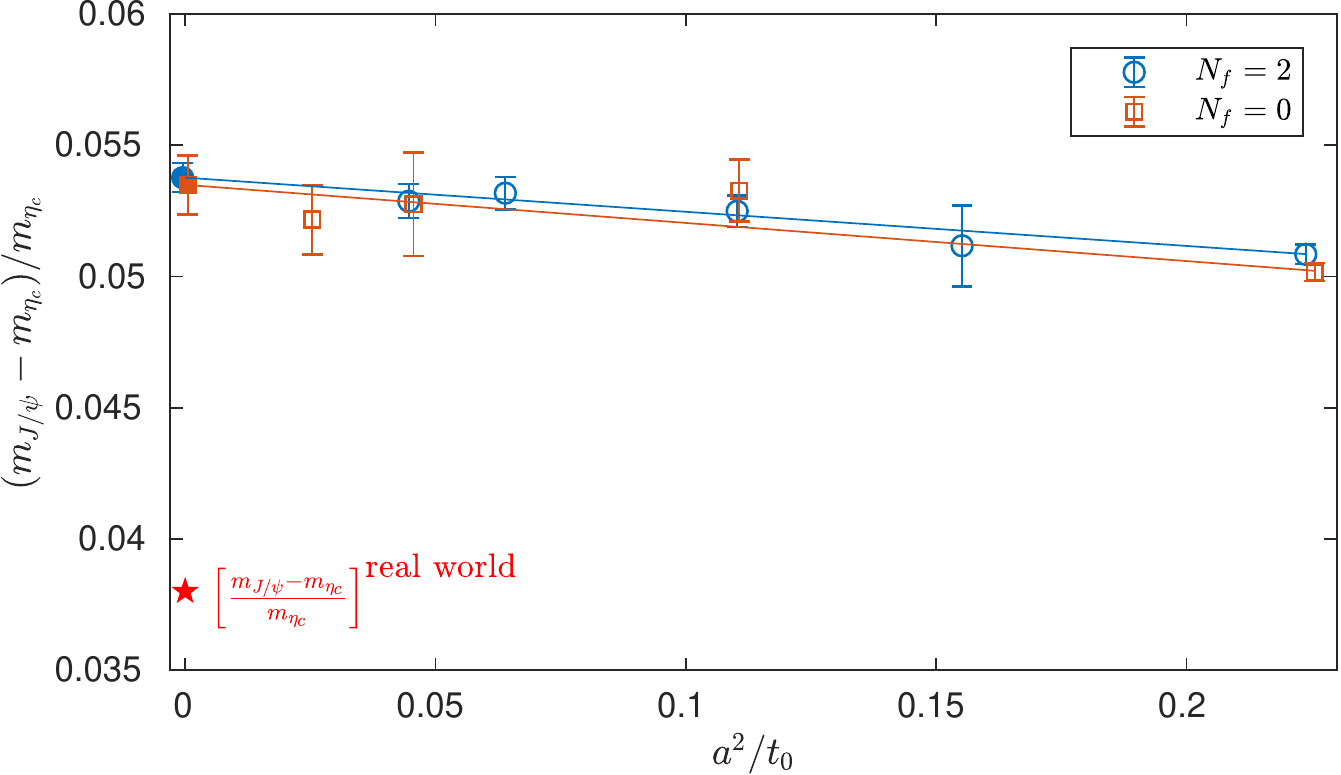}
\caption{\textit{Left panel}: Continuum extrapolation of $m_{J/\psi}$ in $N_f = 0$ and $N_f = 2$ QCD, performed for lattice spacings
$a \lesssim 0.06$ fm. \textit{Right panel}: Continuum extrapolation of the hyperfine splitting $(m_{J/\psi} - m_{\eta_c} )/m_{\eta_c}$ in $N_f = 0$
and $N_f = 2$ QCD. The red star denotes the physical value of the hyperfine splitting.}
\label{fig:mvec_and_splitting}
\end{figure}

\begin{figure}
\includegraphics[width=0.49\textwidth]{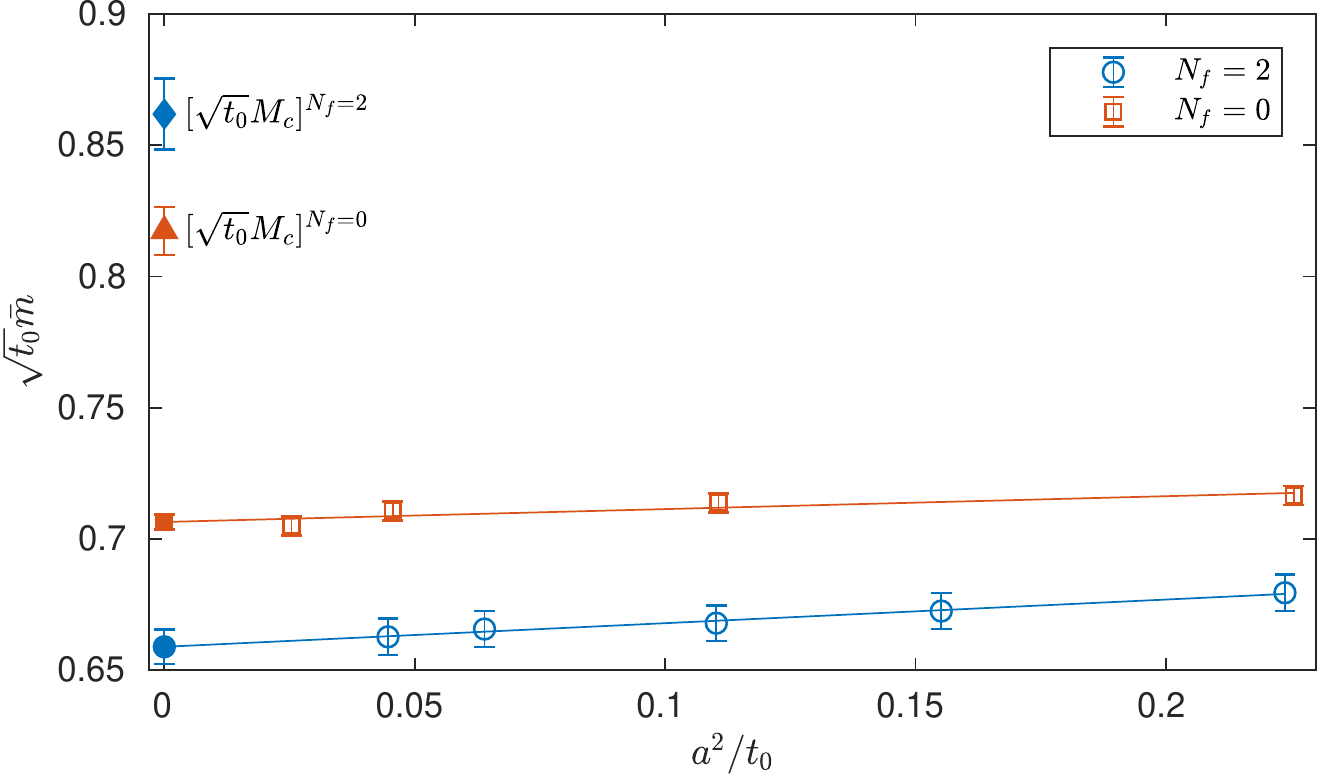}\hfill
\includegraphics[width=0.49\textwidth]{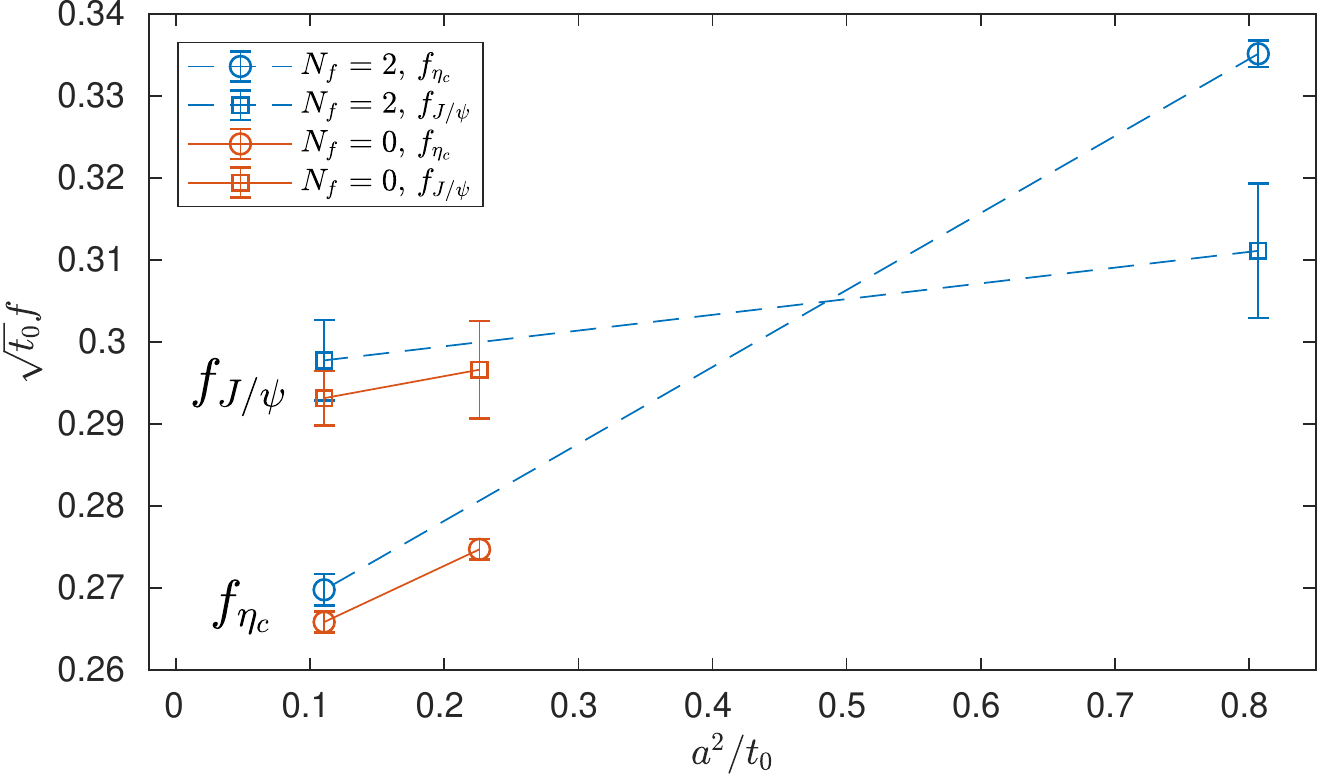}
\caption{\textit{Left panel}: Continuum extrapolation for the running mass $\bar{m}$. The continuum values in the two theories are then converted to the RGI mass $M_c$. \textit{Right panel}: Preliminary results of the meson decay constants $f_{\eta_c}$ and $f_{J/\psi}$ on our $N_f = 0$ (red markers) and $N_f = 2$ (blue markers) ensembles.}
\label{fig:rgimass_and_decay}
\end{figure}

After Wick contractions, the computation of the meson correlators and their derivatives with respect to $\mu$ requires the evaluation of traces of matrices, that we compute making use of stochastic time-diluted estimators with 16 U(1) noise vectors. Our results are summarized in Figs. \ref{fig:eff_mass_and_latt_art}, \ref{fig:mvec_and_splitting} and \ref{fig:rgimass_and_decay}.

In Fig.~\ref{fig:eff_mass_and_latt_art} (left panel) we show the effective masses and the plateaux averages for the $N_f=2$ ensemble with $\beta=6$. As can be seen, our numerical setup allows to take the plateau averages of the effective masses for a large range of temporal slices and this is crucial to determine the size of the charm sea effects with great accuracy. On the right panel of Fig.~\ref{fig:eff_mass_and_latt_art} a study of the lattice artifacts for the vector mass $m_{J/\psi}$ is presented. Beyond $a\approx 0.06 - 0.07$ fm the discretization effects are considerable, as was also already found in the context of the precision computation of the $D_s$ meson decay constant in quenched QCD \cite{Heitger:2008jq}. In particular, we find a non-trivial dependence on the lattice spacing and that lattice spacings $a\lesssim 0.06$ fm have to be employed to obtain reliable continuum extrapolations at $1\%$ precision. In Fig.~\ref{fig:mvec_and_splitting} we compare the continuum limits of $\sqrt{t_0}m_{J/\psi}$ (left panel) and of the hyperfine splitting $(m_{J/\psi} - m_{\eta_c} )/m_{\eta_c}$ (right panel) in $N_f = 0$ and $N_f = 2$ QCD. As can be seen, dynamical charm effects on these observables are not resolvable at a precision of $0.1\%$ and $2\%$ respectively. The discrepancy between our continuum estimate of the hyperfine splitting with its physical value is probably due to effects of light sea quarks, disconnected contributions and electromagnetism that are neglected in this work. Finally, in Fig.~\ref{fig:rgimass_and_decay} we present our results for the RGI quark mass $M_c$ (left panel) and preliminary results for the meson decay constants (right panel). To compute the RGI mass, first we determine  the continuum limit of the running masses $\bar{m}$ in $N_f=0$ and $N_f=2$ QCD and then multiply these values by the ratio $M/\bar{m}$, which is known in both theories. In this case the dynamical charm effects seem relevant and we observe a deviation between the RGI masses of the two theories of around $5\%$, albeit with large statistical uncertainty (the effect is $\simeq 2.7\sigma$). As concerns the meson decay constants, we present a preliminary study for some of our ensembles. We see that $f_{J/\psi}$ is almost $10\%$ larger than $f_{\eta_c}$ and it seems less affected by cut-off effects. In a future work, we plan to increase the statistics and explore more lattice spacings to estimate the charm sea effects on $f_{\eta_c}$ and $f_{J/\psi}$ in the continuum limit.

\acknowledgments
We gratefully acknowledge the Gauss Centre for Supercomputing (GCS) for providing computing time on the supercomputers JURECA, JUQUEEN (at the J\"ulich Supercomputing Centre) and SuperMUC (at the Leibniz Supercomputing Centre). S.C. acknowledges support from the European Union’s Horizon 2020 research and innovation programme
under the Marie Sk\l{}odowska-Curie grant agreement No. 642069. We thank J.~Heitger and K.~Eckert for sharing the implementation of the distance preconditioning method.
\bibliographystyle{JHEP}
\bibliography{biblio}



\end{document}